\documentclass[preprint]{aastex}

\usepackage{lscape}
\usepackage{overpic}
\usepackage{color}
\usepackage{url}
\usepackage{graphicx}

\newcommand{\HI}{{\sc H}\,{\scriptsize{\sc I}}}

\definecolor{deepblue}{rgb}{0.15, 0.4, 0.9}

\slugcomment{Draft v3.10}

\shorttitle{Suzaku observation of Cygnus Cocoon}
\shortauthors{Mizuno, Tanabe et al.}

\begin{document}


\title{{\it Suzaku} Observation of the {\it Fermi} Cygnus Cocoon: 
Search for a Signature of Young Cosmic-Ray Electrons}


\author{
T. Mizuno\altaffilmark{1}, 
T. Tanabe\altaffilmark{1}, 
H. Takahashi\altaffilmark{1},
K. Hayashi\altaffilmark{2},
R. Yamazaki\altaffilmark{3},
I. Grenier\altaffilmark{4},
L. Tibaldo\altaffilmark{5},
}

\email{mizuno@hep01.hepl.hiroshima-u.ac.jp}


\altaffiltext{1}{Hiroshima University}
\altaffiltext{1}{ISAS/JAXA}
\altaffiltext{1}{Aoyama Gakuin University}
\altaffiltext{1}{CEA Sacray}
\altaffiltext{1}{SLAC National Accelerator Laboratory}


\begin{abstract}
The origin of Galactic cosmic rays remains unconfirmed, but promising 
candidates for their sources
are found in star-forming regions. 
We report a series of X-ray observations, with {\it Suzaku}, toward the nearby star-forming region of Cygnus X. They aim at comparing diffuse X-ray emissions on and off the $\gamma$-ray cocoon of hard cosmic rays revealed 
by {\it Fermi} LAT. 
After excluding point sources and small-scale structures and subtracting the non-X-ray and 
cosmic X-ray backgrounds, the 2--10~keV X-ray intensity distribution is found to monotonically decrease 
with increasing Galactic latitude. 
This indicates that most of the extended emission detected by {\it Suzaku} originates from the Galactic ridge.
In two observations, we derive upper limits of $3.4 \times 10^{-8}~{\rm erg~s^{-1}~cm^{-2}~sr^{-1}}$ and
$1.3 \times 10^{-8}~{\rm erg~s^{-1}~cm^{-2}~sr^{-1}}$ to X-ray emission in the 2--10 keV range from the $\gamma$-ray cocoon.
These limits exclude the presence of cosmic-ray electrons with energies above about 50~TeV at a flux level capable of explaining the $\gamma$-ray spectrum. They are consistent with the emission cut-off observed near a TeV in $\gamma$ rays. 
The properties of Galactic-ridge and local diffuse X-rays are also discussed.
\end{abstract}


\keywords{ISM: cosmic-rays --- X-rays: diffuse background --- X-rays: ISM --- gamma rays: general}



\section{Introduction}

An important question of modern astrophysics is the origin and propagation of  
Galactic cosmic rays (GCRs),
which are charged particles with relativistic energies (up to
$10^{15\mbox{--}16}$~eV) diffusing in and around the Milky Way. At low energy, they directly affect the 
chemistry and thermodynamics of interstellar matter through ionization,  
heating, and pressure.
Supernova remnants (SNRs) are widely considered as the most plausible sources of GCRs, 
because they are energetic and numerous enough to maintain the power of 
GCRs 
\citep[e.g.,][]{Ginzburg1964}.
Because the massive OB stars (the progenitors of core-collapse supernovae) 
are born 
in clusters and live shortly, supernova explosions tend to cluster in space (within a few parsecs) and in time (within a few $10^5$ years) \citep{higdon05}.
Therefore, GCRs are expected to be accelerated and injected into the interstellar 
space, not only by the ensembles of individual SNRs, but also by overlapping shocks from SNRs 
and massive stellar winds  \citep[called superbubbles,][]{tenorio88} 
created around OB associations \citep[e.g.,][]{bykov92,parizot04}.
However, this scenario about GCR origin requires 
observational confirmation.

Recently, the Large Area Telescope (LAT) on the {\it Fermi} Gamma-Ray Space Telescope \citep{Atwood2009}
has revealed an extended source of hard, multi-GeV $\gamma$ rays, called the "Cygnus cocoon" \citep{Ackermann2011},
in the nearby star-forming region known as Cygnus X
\citep[e.g.,][]{Piddington1952,Uyaniker2001}, at a distance of $\sim 1.4~{\rm kpc}$ \citep{rygl12}. The extended emission is detected with high confidence above the interstellar $\gamma$-ray background and after subtraction of known point sources, at energies above 1~GeV. The emission morphology corresponds to the region bounded by the ionization fronts powered by the numerous OB stars present in the region. The morphology and the absence of marked spectral variations across the cocoon
imply an interstellar origin rather than a superposition of unresolved $\gamma$-ray sources. 
The observed $\gamma$-ray spectrum cannot be explained by the 
GCR proton or electron spectra measured at the Earth and in the local interstellar medium. The latter is often referred to as the Local Interstellar Spectrum (LIS). An amplification factor of $(1.5\mbox{--}2) \times (E_p/{\rm 10~GeV})^{0.3}$ of the LIS  proton spectrum, or of 
$\sim 60 \times (E_e/{\rm 10~GeV})^{0.5}$ of the LIS electron spectrum,
is required to explain the cocoon emission (where $E_p$ and $E_e$ respectively denote the energy of the CR proton and electron in the cocoon). 
The intense and hard $\gamma$-ray spectrum thus indicates the presence of 
freshly-accelerated CRs in the cocoon.
 
{\it Fermi}-LAT data alone, however, cannot constrain the dominant type of radiating particles (protons 
or electrons), nor the maximum energy of the electrons. For this purpose, 
$\gamma$-ray observations are usefully complemented 
by X-ray observations, which are sensitive to CR electrons at TeV energies. 
We have therefore conducted deep X-ray observations of the $\gamma$-ray cocoon 
region using the X-ray Imaging Spectrometer, XIS \citep{Koyama2007}
on board {\it Suzaku} satellite \citep{Mitsuda2007}, 
which is very sensitive to extended X-ray emission.
If the cocoon $\gamma$-ray emission is produced by interactions of CR electrons with the 
interstellar radiation field and the particles accelerated beyond 10~TeV in energy
have not yet cooled or escaped the region,
we expect to find extended non-thermal X-ray emission due to synchrotron radiation 
in the large magnetic field of ${\rm \sim 20~\mu G}$ 
inferred in the region  \citep{Ackermann2011}.

\section{Observations and Data Reduction}


To constrain the CR properties in the cocoon, 
we have performed a series of X-ray observations of the Cygnus region. 
Two ``source'' observations were pointed at positions with strong GeV 
cocoon emission, 
but free of known, bright X-ray sources, such as the SNR $\gamma$~Cygni and the 
binary Cygnus X-3. 
Because a potential X-ray counterpart to the cocoon is likely to fill the whole field of view of the XIS instrument,
we have also conducted two ``background'' observations to
estimate the backgrounds in the Cygnus direction, in
particular the intensity of the Galactic ridge X-ray emission 
\citep[GRXE;][]{Worrall1982,Warwick1985,Koyama1986}.
The GRXE is an apparently diffuse feature along the Galactic plane
and its origin would be a mixture of truly diffuse hot plasma and numerous dim point sources
\citep[e.g.,][]{Ebisawa2005,Revnivtsev2006,Uchiyama2013}.
The positions of the four observations are overlaid on the $\gamma$-ray  
count map of the cocoon in Figure~1.

The observations were carried out using the XIS
on the focal plane of the X-Ray Telescope \citep[XRT:][]{Serlemitsos2007} on {\it Suzaku}. 
The XIS consists of two front-illuminated (FI) X-ray CCDs (XIS0 and 3) 
\footnote{Because of an anomaly occurred in 2006 November, the operation of 
another FI sensor, XIS2, has been terminated.}
and one backside-illuminated (BI) X-ray CCD (XIS1). 
The combined XIS and XRT system is sensitive within the energy range 
0.3--12~keV. 
Although its angular resolution is moderate (half-power diameter $\sim 2'$),
the XIS+XRT system provides a low and stable instrumental background 
\citep{Mitsuda2007,Tawa2008}; therefore it is suitable for the study of extended
emission
with low surface brightness. 
The XIS was operated in the normal clocking full-frame mode, and data were recorded in the 
$3 \times 3$ or $5 \times 5$ editing mode. Data were analyzed using the HEADAS 
6.15.1 software with the calibration database
released on July 1, 2014. We reprocessed data using {\tt aepipeline} to take the latest calibration into account.
We analyzed the so-called cleaned events, which had passed the following standard event selection criteria: 
a) only {\it ASCA}-grade 0, 2, 3, 4, and 6 events were accumulated with hot and
flickering pixels removed;
b) more than 436~s had elapsed since passing through the South Atlantic 
Anomaly; and 
c) the pointing direction was at least $5^{\circ}$ and $20^{\circ}$ above the rim of the Earth 
during night and day, respectively. 
To further reduce the non-X-ray background (NXB), we also required that 
d) the geomagnetic cutoff rigidity exceeded 6~GV. 
Details concerning the observation and net exposures of the screened 
events are summarized in Table~1.

\begin{figure}[htbp]
\centering
\includegraphics[width=0.34\textwidth]{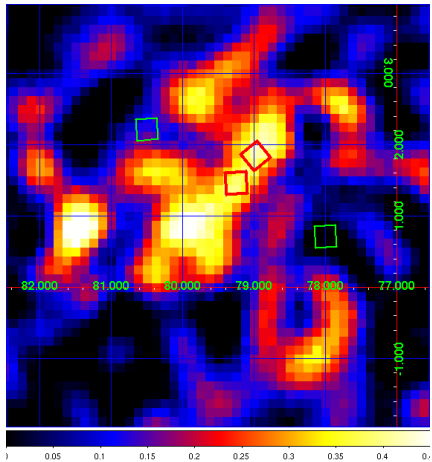}
\caption{Positions of X-ray observations overlaid on the $\gamma$-ray residual count map of the cocoon above 10~GeV, smoothed with a $\sigma=0.\hspace{-2pt}^{\circ}25$ Gaussian kernel
\citep[][]{Ackermann2011}. The unit of the $\gamma$-ray map is counts per pixel after
the smoothing was applied.
The coordinates are in Galactic longitude and latitude.
Source and background observations are indicated by thick red and thin green boxes, respectively.
See Table~1 for details.}
\end{figure}

\begin{table}[htbp]
\caption{Summary of observations}
\begin{center}
\small
\begin{tabular}{ccccccc} \hline\hline
Region & \multicolumn{2}{c}{Pointing\tablenotemark{a}} & Observation date & Net exposure  \\ 
       & l (deg)          & b (deg)   &           & (ks)  \\   \hline
Source 1     & 79.25 & 1.49 & 2012 Nov 18 & 43.3 \\
Source 2     & 78.99 & 1.86 & 2013 Dec 07 & 42.0  \\
Background 1 & 78.00 & 0.74 & 2012 Nov 19 & 19.9  \\
Background 2 & 80.50 & 2.24 & 2012 Nov 19 & 25.6  \\ \hline 
\end{tabular}
\tablenotetext{a}{$l$ and $b$ are Galactic longitude and latitude of the center of the XIS field-of-view, respectively.}
\end{center}
\end{table}

\clearpage




\section{Data Analysis and Results}

\subsection{X-ray Images}

We have extracted X-ray images from the two FI CCDs (XIS0 and XIS3). 
These cameras have better imaging quality than the BI CCD (XIS1) due to their 
lower instrumental background. 
We have defined the soft and hard bands as 0.7--2 keV and 2--10 keV, respectively, 
and have excluded the corners of the CCD chips illuminated by the ${\rm ^{55}Fe}$ calibration sources. 
We have then estimated the NXB contribution from the night Earth data and subtracted it from the images using 
{\tt xisnxbgen} \citep{Tawa2008}.
Vignetting was then corrected by dividing the soft- and hard-band images by flat sky images simulated at 
1~keV and 4~keV, respectively, by using the XRT+XIS simulator {\tt xissim} \citep{Ishisaki2007}. 
In the flat image simulations, we have assumed a uniform intensity of ${\rm 1~photon~s^{-1}~cm^{-2}~sr^{-1}}$,
thus the approximate unit of the obtained vignetting-corrected image is 
${\rm photons~s^{-1}~cm^{-2}~sr^{-1}}$. 
We have combined two FI CCD images and applied smoothing with a Gaussian kernel of $\sigma=0.\hspace{-2pt}'28$ for visualization.
Several point sources and small-scale structures are recognizable in the final images (see Figure~2). 
Moreover, we also observe strong, apparently extended emission in the soft-band 
images of Source~2 and Background~2 observations. In the hard-band images, the 
apparently extended emission is more intense 
in Source~1 and Background~1 than in the other two regions.

\begin{figure}[t]
\centering
\begin{overpic}
[width=0.4\textwidth]{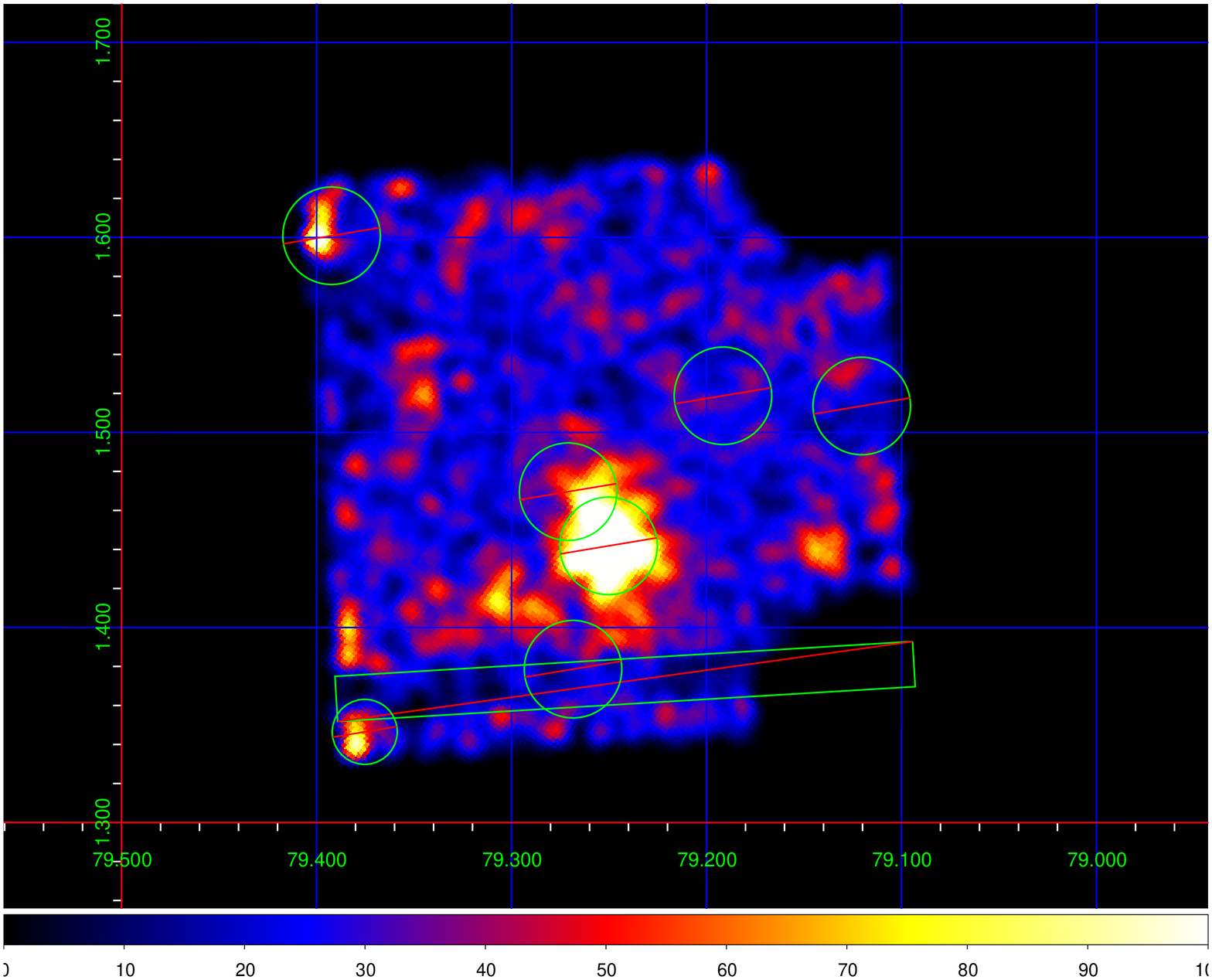}
\put(88,75){\textcolor{white}{(a1)}}
\end{overpic}
\begin{overpic}
[width=0.4\textwidth]{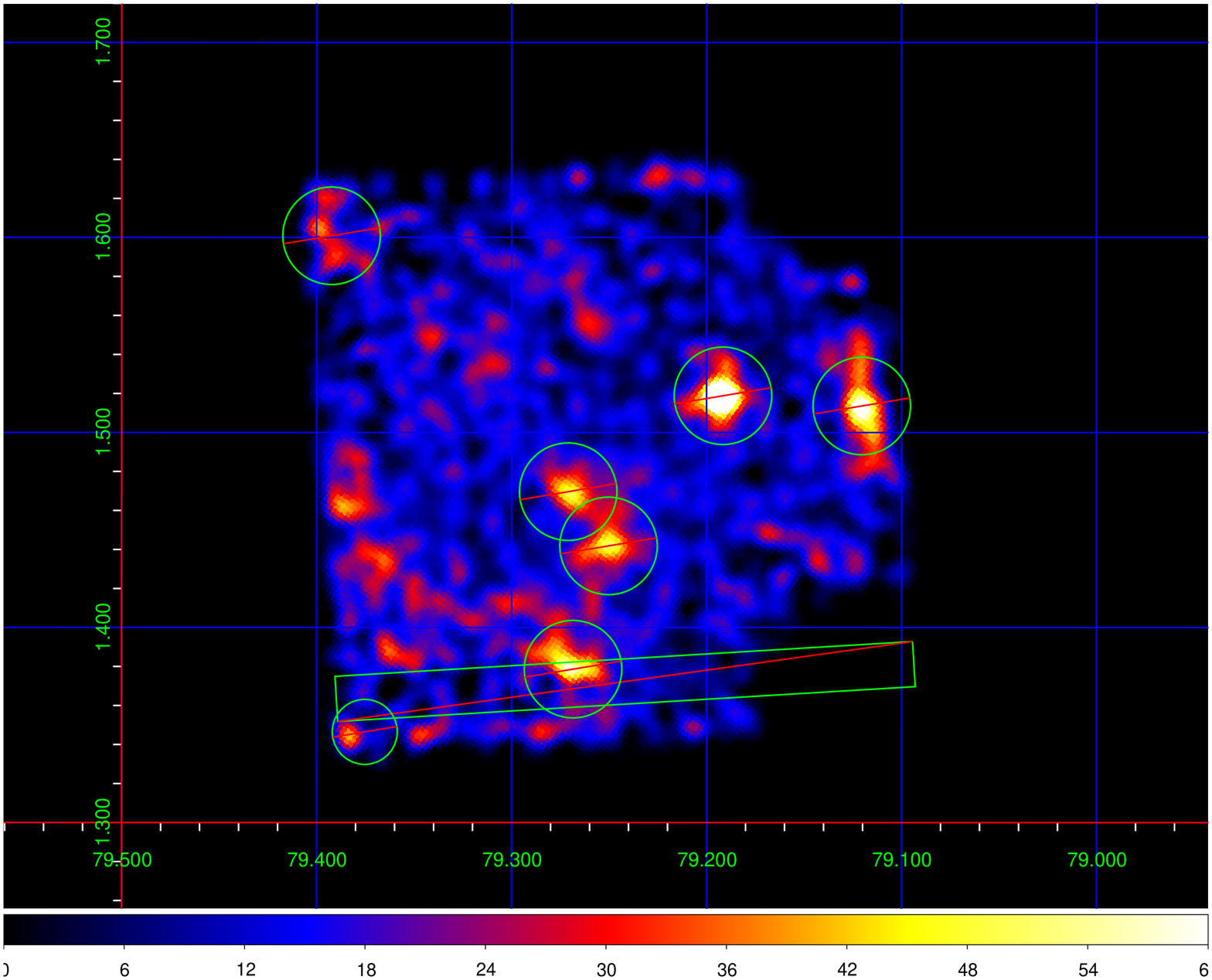}
\put(88,75){\textcolor{white}{(a2)}}
\put(55,35){\textcolor{white}{\footnotesize src1}}
\put(53,55){\textcolor{white}{\footnotesize src2}}
\put(43,47){\textcolor{white}{\footnotesize src3}}
\put(53,27){\textcolor{white}{\footnotesize src4}}
\put(65,54){\textcolor{white}{\footnotesize src5}}
\put(33,60){\textcolor{white}{\footnotesize src6}}
\end{overpic}
\begin{overpic}
[width=0.4\textwidth]{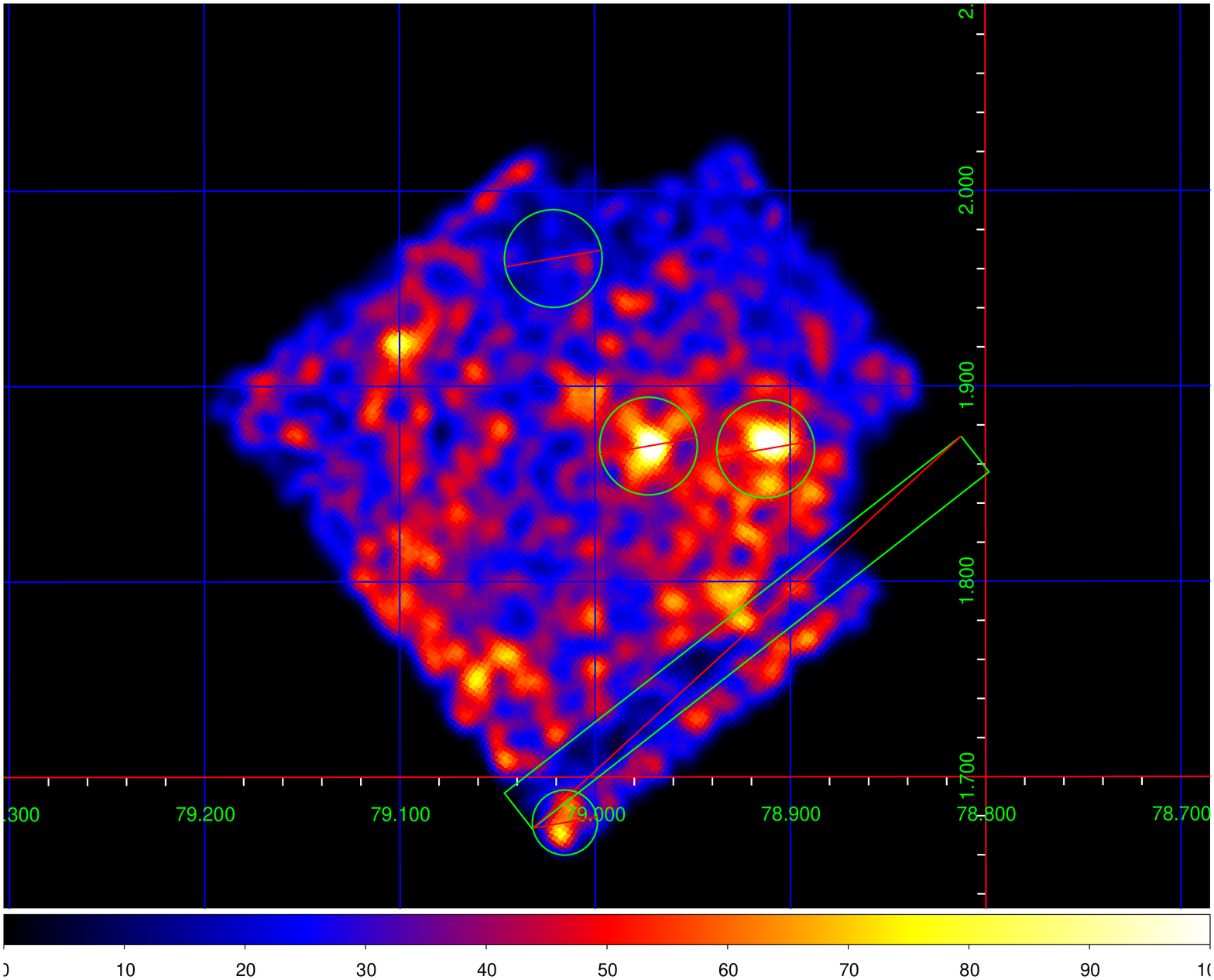}
\put(88,75){\textcolor{white}{(b1)}}
\end{overpic}
\begin{overpic}
[width=0.4\textwidth]{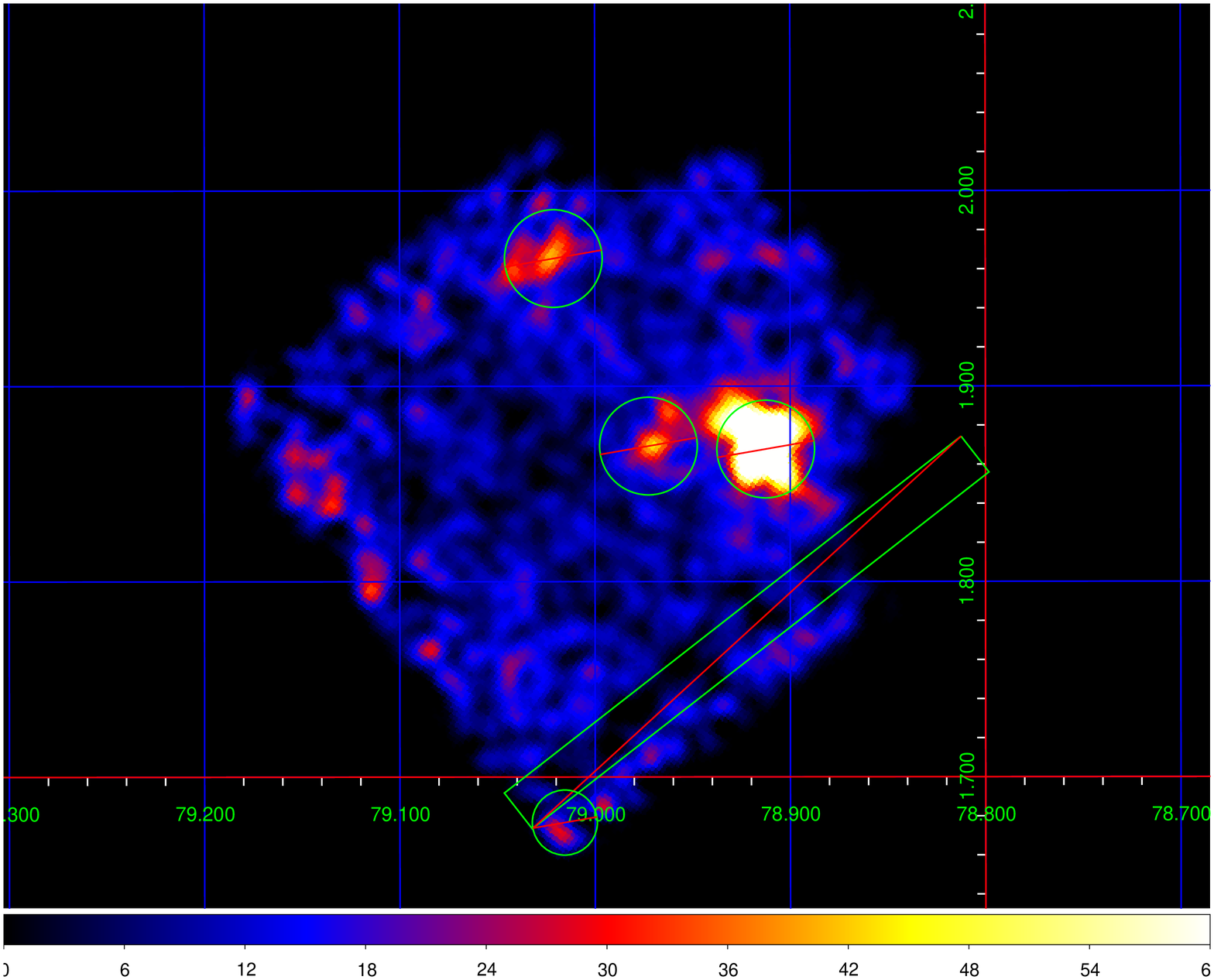}
\put(88,75){\textcolor{white}{(b2)}}
\put(60,50){\textcolor{white}{\footnotesize src1}}
\put(51,60){\textcolor{white}{\footnotesize src2}}
\put(48,50){\textcolor{white}{\footnotesize src3}}
\end{overpic}
\begin{overpic}
[width=0.4\textwidth]{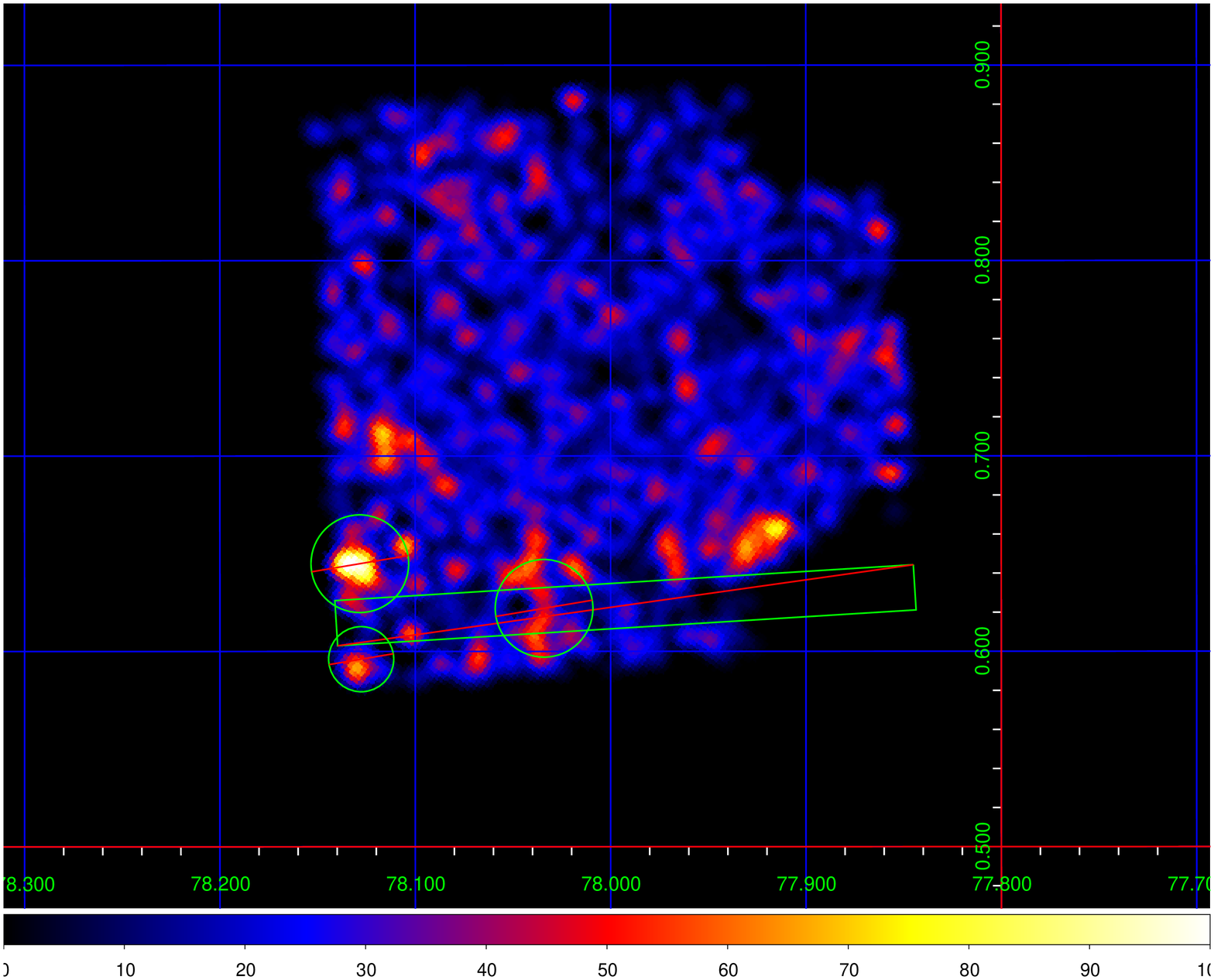}
\put(88,75){\textcolor{white}{(c1)}}
\end{overpic}
\begin{overpic}
[width=0.4\textwidth]{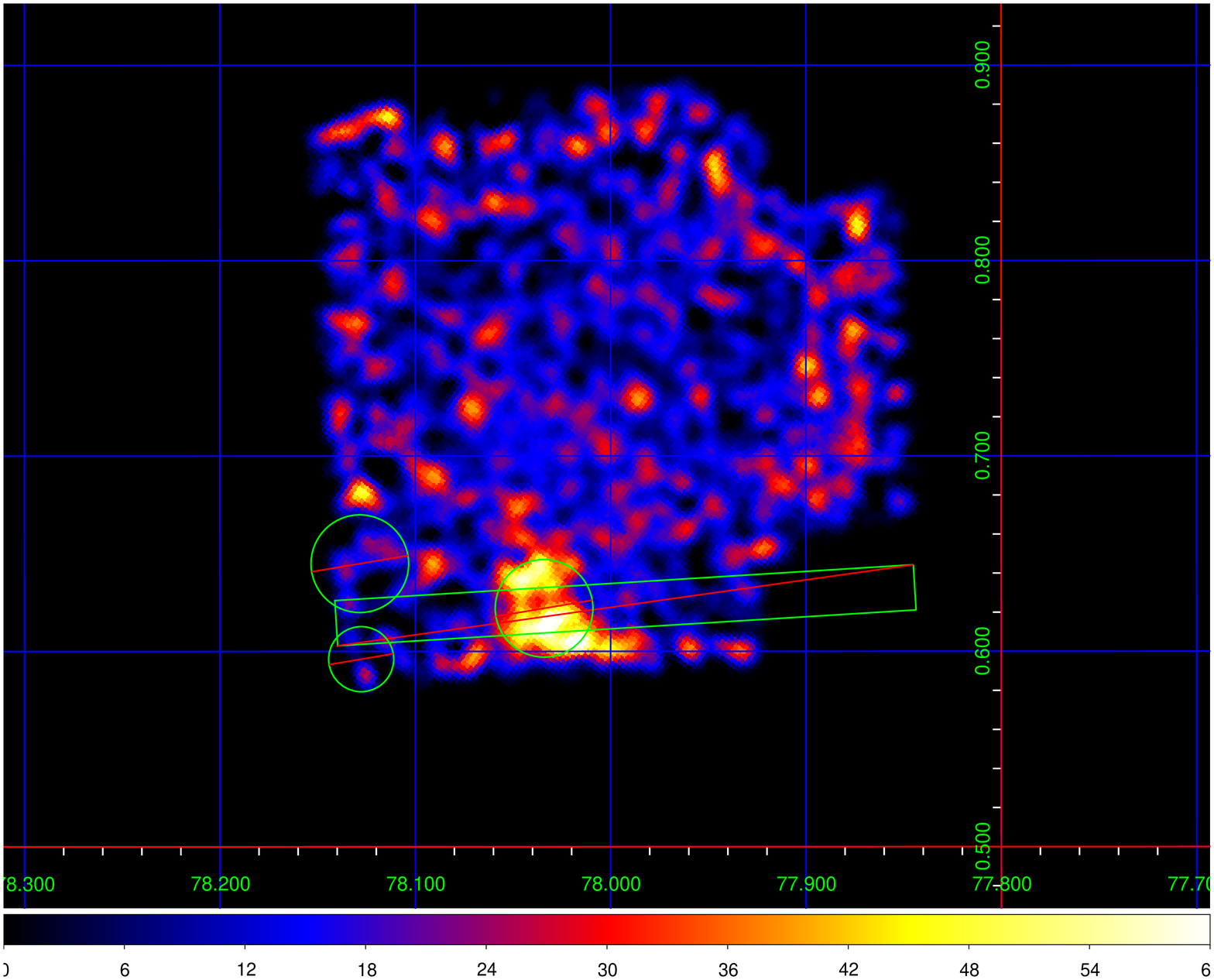}
\put(88,75){\textcolor{white}{(c2)}}
\put(45,37){\textcolor{white}{\footnotesize src1}}
\put(32,38){\textcolor{white}{\footnotesize src2}}
\end{overpic}
\begin{overpic}
[width=0.4\textwidth]{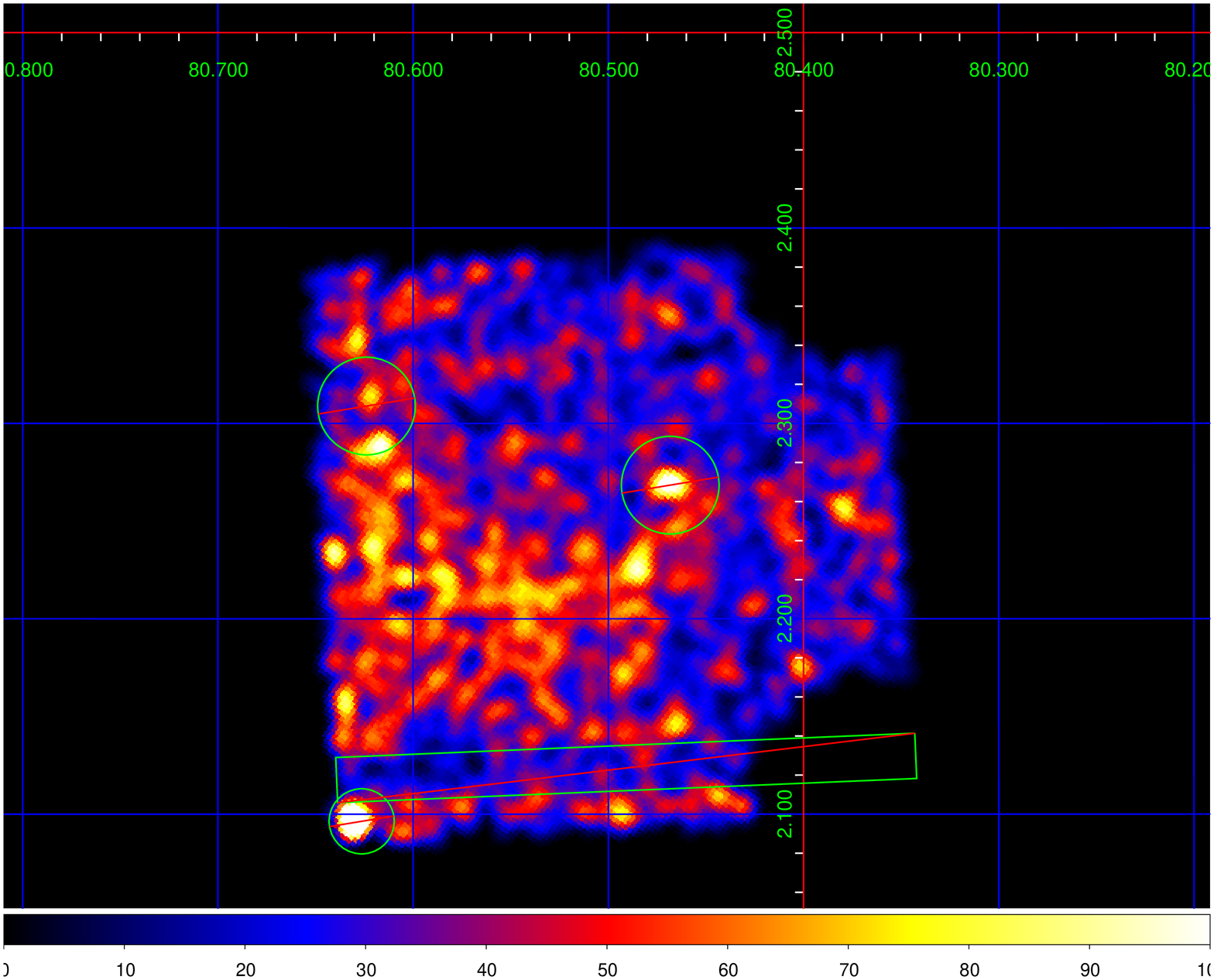}
\put(88,75){\textcolor{white}{(d1)}}
\end{overpic}
\begin{overpic}
[width=0.4\textwidth]{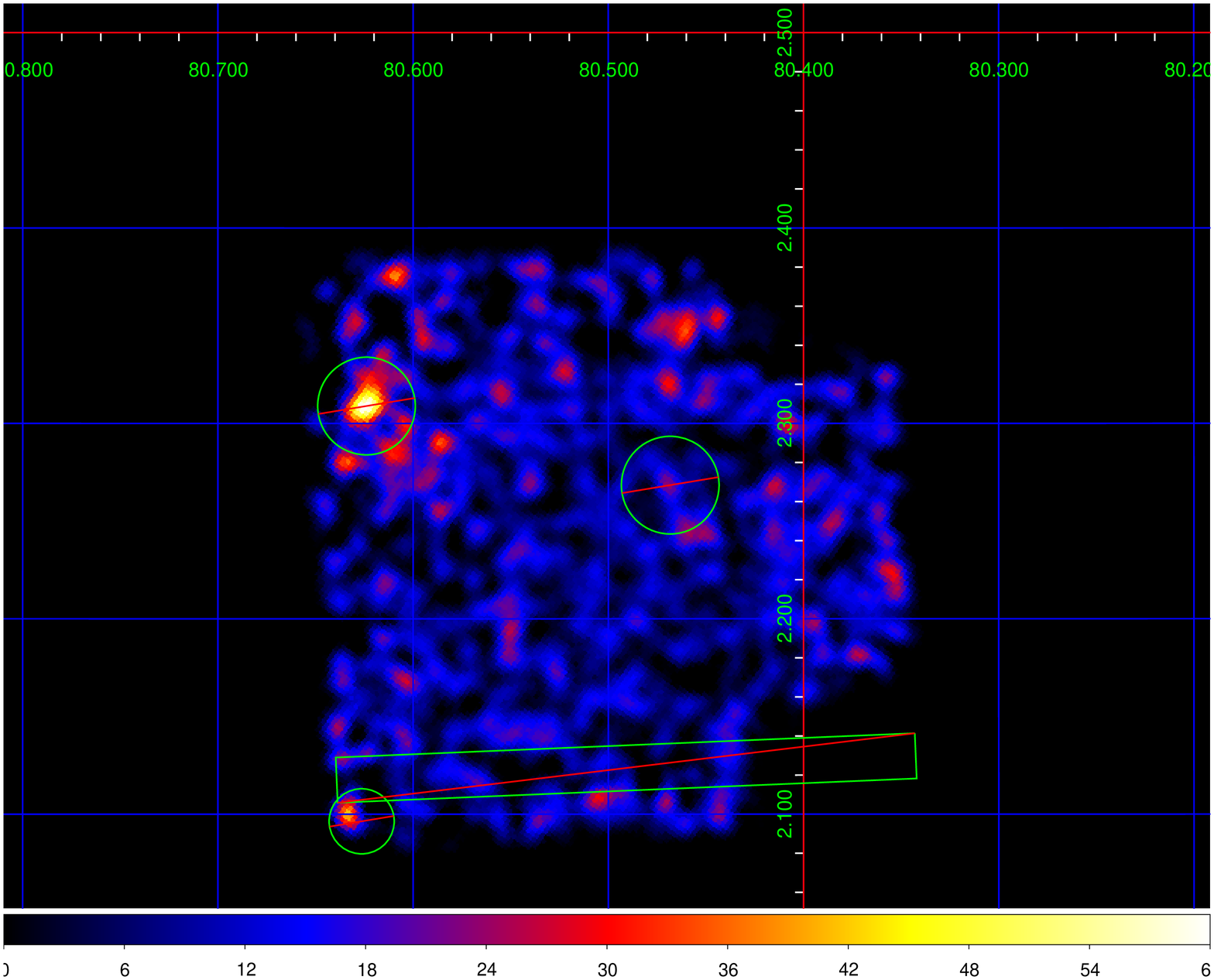}
\put(88,75){\textcolor{white}{(d2)}}
\put(35,45){\textcolor{white}{\footnotesize src1}}
\put(50,30){\textcolor{white}{\footnotesize src2}}
\end{overpic}
\caption{From top to bottom: X-ray intensity maps in Galactic coordinates of Source~1 observation (panels a), 
Source~2 observation (panels b), 
Background~1 observation (panels c), and Background~2 observation (panels d).
Left and right panels show the 
soft- (0.7--2~keV) and hard-band (2--10~keV) images, respectively.}
\end{figure}

\subsection{Extended Emission Spectra}

We have extracted extended emission based on the 
vignetting-corrected images (Figure~2). 
We have first identified point sources and small-scale structures in each of the soft- and hard-band images 
by eye and have delineated them by $1.\hspace{-2pt}'5$ radius circular regions (green circles in Figure~2). 
These structures have been designated as src1, src2, etc. in each field,
and events in those circles have been removed\footnote
{
Although the choice of sources and structures is not unique, the effect on the estimated
extended emission is small. For example, even if we included src3 
(a moderately bright source located close to the center of the field of view)
of Source~1 observation,
the 2--10~keV intensity of the extended emission was increased by $\sim 0.3 \times 10^{-8}~{\rm erg~s^{-1}~cm^{-2}~sr^{-1}}$
which is less than half of the uncertainty examined in \S~4.1 and shown in Figure~5
}.
Spectral analysis of these sources and structures is described in \S~3.2 and results are
summarized in Table~3.
An ${\rm 80~pixel \times 1024~pixel}$ rectangular region and an area 
with (occasionally) significantly high background at the corner of the chip are unusable in 
X1S1\footnote{
See "The Suzaku Data Reduction Guide" of v5.0 (\url{http://www.heasarc.gsfc.nasa.gov/docs/suzaku/analysis/abc})
}.
For simplicity, these regions have been excluded from the sky projections for all XIS camera images. 
We have then accumulated the extended X-ray emission spectra by integrating the events over the whole XIS field of view ($\sim 17.\hspace{-2pt}'8 \times 17.\hspace{-2pt}'8$), and analyzed the spectra 
for diffuse X-ray emission with ancillary response files (ARFs)
calculated by {\tt xissimarfgen} \citep{Ishisaki2007}, assuming uniform brightness 
over a $20'$ radius. 
The loss of effective area due to the exclusion of point sources and unusable area 
described above has been taken into account in calculating the ARFs.
The response matrix files (RMFs) has been calculated with {\tt xisrmfgen}, and the NXB spectrum 
integrated over the same region as that for the extended emission
has been estimated by {\tt xisnxbgen} \citep{Tawa2008}
and subtracted from the source spectrum.
In our observations, the energy resolution degraded by particle radiation was restored 
by a spaced-row charge-injection technique \citep[e.g.,][]{Uchiyama2009a} with a 6~keV equivalent charge applied on orbit. The amount of the charge injected affects the NXB level of the XIS,
and was taken into account in calculationg the NXB.

The NXB-subtracted X-ray spectrum is expected to be a sum of the cosmic X-ray background (CXB), 
the GRXE (plus local diffuse X-ray emission), and the possible X-rays from the 
cocoon. Following 
\cite{Kushino2002},
the CXB has been modelled as a fixed power law with a photon index of 1.4 and a normalization of 
$9.42 \times 10^{-4}~{\rm photons~s^{-1}~cm^{-2}~keV^{-1}}$ at 1~keV,
or a 2--10~keV flux of $6.20 \times 10^{-12}~{\rm erg~s^{-1}~cm^{-2}}$, corresponding to
an intensity of $5.85 \times 10^{-8}~{\rm erg~s^{-1}~cm^{-2}~sr^{-1}}$ 
integrated throughout the $20'$ radius. Adopting the optically-thin 
approximation, from
\citet{Kalberla2005} we have estimated the neutral hydrogen column density toward 
the regions studied here as $\sim 1.3 \times 10^{22}~{\rm cm^{-2}}$.
However, the opacity of the \HI\ 21~cm line is non-negligible in the Galactic plane \citep[e.g.,][]{Paper2},
and contribution from molecular clouds and dark neutral medium \citep[e.g.,][]{DNM}
should be taken into account. We therefore estimated total neutral hydrogen column density
from $\gamma$-ray data in these regions based on the analysis by \citet{CygISM}.
We adopted spin temperature of 250~K as a baseline model and obtained the total
column density $N(\mathrm H)$ of $(2.5\mbox{--}3.1) \times 10^{22}~{\rm cm^{-2}}$, giving the
2--10~keV intensity of $(4.8\mbox{--}5.0) \times 10^{-8}~{\rm erg~s^{-1}~cm^{-2}~sr^{-1}}$
(see Table~2).
The uncertainty due to the assumed spin temperature is evaluated in \S~4.1.
Based on previous
studies of the GRXE and local diffuse X-ray emission
\citep[e.g.,][]{Kaneda1997,Uchiyama2009b,Kataoka2013}, we modeled the GRXE 
(plus local diffuse X-ray emission) as
${\tt vapec1} + {\tt wabs2} \times {\tt apec2} + {\tt wabs3} \times {\tt apec3}$ in {\tt XSPEC},
where {\tt apec} (and {\tt vapec}) and {\tt wabs} are thin-thermal plasma models
by \citet{Smith2001} and interstellar absorption by \citet{Morrison1983}, respectively.
We have allowed the relative normalizations among the three XIS sensors to vary.
Among three components, the first term (low-temperature plasma model) 
represents the unabsorbed Local Bubble emission and/or contamination
from the Solar-Wind Charge eXchange (SWCX) \citep[e.g.,][]{Fujimoto2007}, and we have fixed the
plasma temperature and abundance to $kT=0.1~{\rm keV}$ and $Z=Z_{\sun}$, respectively
\citep[see, e.g.,][]{Smith2007,Yoshino2009}. The second and the third terms 
(mid-temperature and high-temperature plasma models) correspond to the
classical soft-temperature and hard-temperature emissions of the GRXE, respectively \citep[e.g.,][]{Kaneda1997}.
In all four observations, we have found that our low-temperature plasma model is not able to reproduce the data
in the energy range of 0.5--0.6~keV, because the observed energy of the emission line
is lower than the {{\sc O}\,{\scriptsize{\sc VII}}}
resonance line energy
(0.574~keV). We have therefore set the Oxygen abundance to 0 and added a gaussian line at $\sim 0.56~{\rm keV}$
with the line center energy and normalization allowed to vary.
The line is likely to be due to a charge-exchange process,
as described in \S~4.2.
We have also added lines at $\sim 0.65~{\rm keV}$ (to reproduce the line of {{\sc O}\,{\scriptsize{\sc VIII}}})
and at $6.40~{\rm keV}$ (to reproduce Fe ${\rm K}_{\alpha}$ line) if the
presence of the line was significant at the 99\% confidence from an F-test. 
Our CXB+GRXE (plus local diffuse X-ray emission) model successfully reproduces the 
Source~1 observation data 
(reduced chi-square $\chi^{2}/{\rm d.o.f.} = 538.1/501$). 
As shown in Figure~3, low-, middle- and high-temperature plasma models dominate the spectrum
below ${\rm \sim 0.7~keV}$, from $\sim 0.7$ to $\sim 2$ keV and above $\sim$ 2~keV, respectively.
Above 2~keV, the contributions from the CXB (fixed to a model described above)
and the high-temperature plasma model are comparable. 
These two components are degenerate to a large extent in spectral shape,
and the effect of the CXB uncertainty on the remaining hard-band intensity is described in \S~4.1.
The results from the fits in the four regions are summarized in Table~2 and 
Figure~4.

\begin{table}
\begin{center}
\caption{Best-fit spectral parameters of extended emission}
\begin{tabular}{ccccc}
 \tableline\tableline
 & Source1 & Source2 & Background1 & Background2 \\ \tableline
$E_{1}$(keV) & $0.548 \pm 0.004$ & $0.545 \pm 0.006$ & $0.551 \pm 0.007$ & $0.554^{+0.009}_{-0.010}$ \\
${\rm Norm_{1}}$ & $23.8^{+3.7}_{-3.9}$ & $24.4 \pm 3.1$ & $22.1^{+5.4}_{-4.7}$ & $14.9^{+5.3}_{-4.0}$ \\
$E_{2}$(keV) & 0.653(fixed)\tablenotemark{a} & $0.648^{+0.013}_{-0.014}$ & $0.673^{+0.031}_{-0.035}$ & -- \\
${\rm Norm_{2}}$ & $2.3^{+0.8}_{-0.9}$ & $5.3 \pm 1.1 $ & $2.0^{+0.9}_{-0.8}$ & -- \\
$E_{3}$(keV) & 6.40(fixed) & -- & -- & -- \\
${\rm Norm_{3}}$ & $0.12^{+0.09}_{-0.10}$ & -- & -- & -- \\
$N({\mathrm{H})}_{\rm low}$($10^{22}~{\rm cm^{-2}}$) & 0(fixed) & 0(fixed) & 0(fixed) & 0(fixed) \\
$kT_{\rm low}$(keV) & 0.1(fixed) & 0.1(fixed) & 0.1(fixed) & 0.1(fixed) \\
$A_{\rm low}$($Z_{\sun}$) & 1(fixed) & 1(fixed) & 1(fixed) & 1(fixed) \\
$EM_{\rm low}$ & $103 \pm 18$ & $144^{+21}_{-20}$ & $99^{+23}_{-24}$ & $139^{+36}_{-34}$ \\
$N({\mathrm{H})}_{\rm mid}$($10^{22}~{\rm cm^{-2}}$) & $0.39^{+0.07}_{-0.09}$ & $0.46^{+0.04}_{-0.06}$ & $0.73^{+0.11}_{-0.14}$ & $0(\le0.02)$ \\
$kT_{\rm mid}$(keV) & $0.73^{+0.02}_{-0.03}$ & $0.62\pm0.02$ & $0.62^{+0.11}_{-0.04}$ & $0.60^{+0.02}_{-0.01}$ \\
$A_{\rm mid}$($Z_{\sun}$) & $0.13^{+0.06}_{-0.04}$ & $0.29^{+0.27}_{-0.10}$ & $0.27^{+1.25}_{-0.15}$ & $0.37^{+1.22}_{-0.11}$ \\
$EM_{\rm mid}$ & $256^{+31}_{-49}$ & $242^{+64}_{-98}$ & $(20^{+9}_{-16})\times10$ & $73^{+26}_{-31}$ \\
$N({\mathrm{H})}_{\rm high}$($10^{22}~{\rm cm^{-2}}$) & $3.1^{+1.4}_{-1.5}$ & $1.9(\le 3.0)\tablenotemark{b}$ & $2.6^{+0.8}_{-1.0}$ & $2.0^{+0.7}_{-1.4}$ \\
$kT_{\rm high}$(keV) & $3.1^{+2.4}_{-0.9}$ & $2.6^{+1.0}_{-0.6}$ & $2.4^{+0.8}_{-0.4}$ & $1.5^{+1.7}_{-0.3}$ \\
$A_{\rm high}$($Z_{\sun}$) & $0.36^{+0.22}_{-0.17}$ & $0.28^{+0.31}_{-0.26}$ & $0.19^{+0.16}_{-0.14}$ & $0.4(\le1.1)$ \\
$EM_{\rm high}$ & $106^{+59}_{-43}$ & $76^{+39}_{-23}$ & $217^{+78}_{-65}$ & $103^{+65}_{-37}$ \\
$I_{\rm low}$(0.5--10~keV)\tablenotemark{c} & $2.46 \times 10^{-8}$ & $2.86 \times 10^{-8}$ & $2.38 \times 10^{-8}$ & $1.55 \times 10^{-8}$ \\ 
$I_{\rm mid}$(0.5--10~keV)\tablenotemark{c} & $5.66 \times 10^{-8}$ & $6.38 \times 10^{-8}$ & $3.01 \times 10^{-8}$ & $8.42 \times 10^{-8}$ \\ 
$I_{\rm high}$(0.5--10~keV)\tablenotemark{c} & $4.88 \times 10^{-8}$ & $3.15 \times 10^{-8}$ & $7.20 \times 10^{-8}$ & $2.39 \times 10^{-8}$ \\ 
$N({\mathrm{H})}_{\rm CXB}$($10^{22}~{\rm cm^{-2}}$) & 2.92(fixed) & 2.47(fixed) & 3.07(fixed) & 2.67(fixed) \\ 
$I_{\rm CXB}$(0.5--2~keV)\tablenotemark{c} & $0.12 \times 10^{-8}$(fixed) & $0.16 \times 10^{-8}$(fixed) & $0.11 \times 10^{-8}$(fixed) & $0.14 \times 10^{-8}$(fixed) \\ 
$I_{\rm CXB}$(2--10~keV)\tablenotemark{c} & $4.85 \times 10^{-8}$(fixed) & $4.97 \times 10^{-8}$(fixed) & $4.81 \times 10^{-8}$(fixed) & $4.92 \times 10^{-8}$(fixed) \\ 
$I_{0.5\mbox{--}2}$\tablenotemark{c} & $7.69 \times 10^{-8}$ & $9.22 \times 10^{-8}$ & $5.72 \times 10^{-8}$ & $10.44 \times 10^{-8}$ \\ 
$I_{2\mbox{--}10}$\tablenotemark{c} & $10.29 \times 10^{-8}$ & $8.33 \times 10^{-8}$ & $11.79 \times 10^{-8}$ & $6.98 \times 10^{-8}$ \\ 
$\chi^{2}$/d.o.f. & 538.1/501 & 565.7/521 & 320.1/252 & 370.5/336 \\ 
\tableline
\end{tabular}
\tablenotetext{}{
$E$ and Norm are line center energy (keV) and the intensity (${\rm photons~s^{-1}~cm^{-2}~sr^{-1}}$) of gaussian, respectively.
$N({\mathrm H})_{\rm low}/kT_{\rm low}/A_{\rm low}/EM_{\rm low}$, $N({\mathrm H})_{\rm mid}/kT_{\rm mid}/A_{\rm mid}/EM_{\rm mid}$ and $N({\mathrm H})_{\rm high}/kT_{\rm high}/A_{\rm high}/EM_{\rm high}$ are absorption/temperature/abundance/emission measure of the
low-, middle- and high-temperature plasma models, respectively. 
The emission measures are given as the value integrated over the line-of-sight, 
$\frac{1}{4\pi}\int n_{\mathrm e} n_{\mathrm H} ds$ 
(where $n_{\mathrm e}$ and $n_{\mathrm H}$ are the electron and hydrogen density, respectively)
in the unit of ${\rm 10^{14}~cm^{-5}~sr^{-1}}$.
$I_{\rm low}$, $I_{\rm mid}$ and $I_{\rm high}$ are absorption-uncorrected intensity of
low-, middle- and high-temperature components, respectively, where associated line fluxes are also included.
$N({\mathrm H})_{\rm CXB}$ and $I_{\rm CXB}$ are
the assumed column density and absorption-uncorrected intensity of the CXB model, respectively.
$I_{0.5\mbox{--}2}$ and $I_{2\mbox{--}10}$ are intensity in 0.5--2 and 2--10~keV including CXB contribution,
respectively.
}
\tablenotetext{a}{The line center energy was not well determined and thus fixed at the energy of
{{\sc O}\,{\scriptsize{\sc VIII}}} ${\rm K}_{\alpha}$ line.}
\tablenotetext{b}{The Lower limit was set at ${\rm 1.0 \times 10^{22}~cm^{-2}}$, otherwise the fit gives unphysical results.
($A_{\rm mid} \sim 5$, $A_{\rm high} \sim 0$ and $N({\mathrm{H}})_{\rm high} \sim 0$)}
\tablenotetext{c}{The unit is given in ${\rm erg~s^{-1}~cm^{-2}~sr^{-1}}$.}
\end{center}
\end{table}

\begin{figure}[htbp]
\centering
\includegraphics[height=0.45\textwidth,angle=-90]{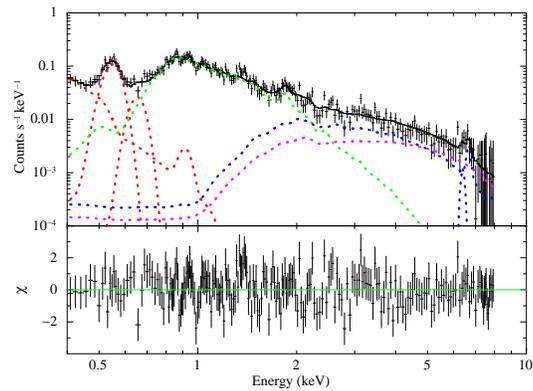}
\caption{
Data and modeled spectra of extended X-ray emission of Source~1 region, taken by XIS1. 
The spectrum is fitted by three-temperature thin-thermal plasmas (GRXE plus local diffuse X-rays) 
and a fixed power law (CXB). 
Low-, middle-, and high-temperature models are indicated by red, green, and blue dotted lines, respectively,
and the CXB contribution is shown by the dotted purple line. The bottom panel shows the residuals.
}
\end{figure}

\begin{figure}[htbp]
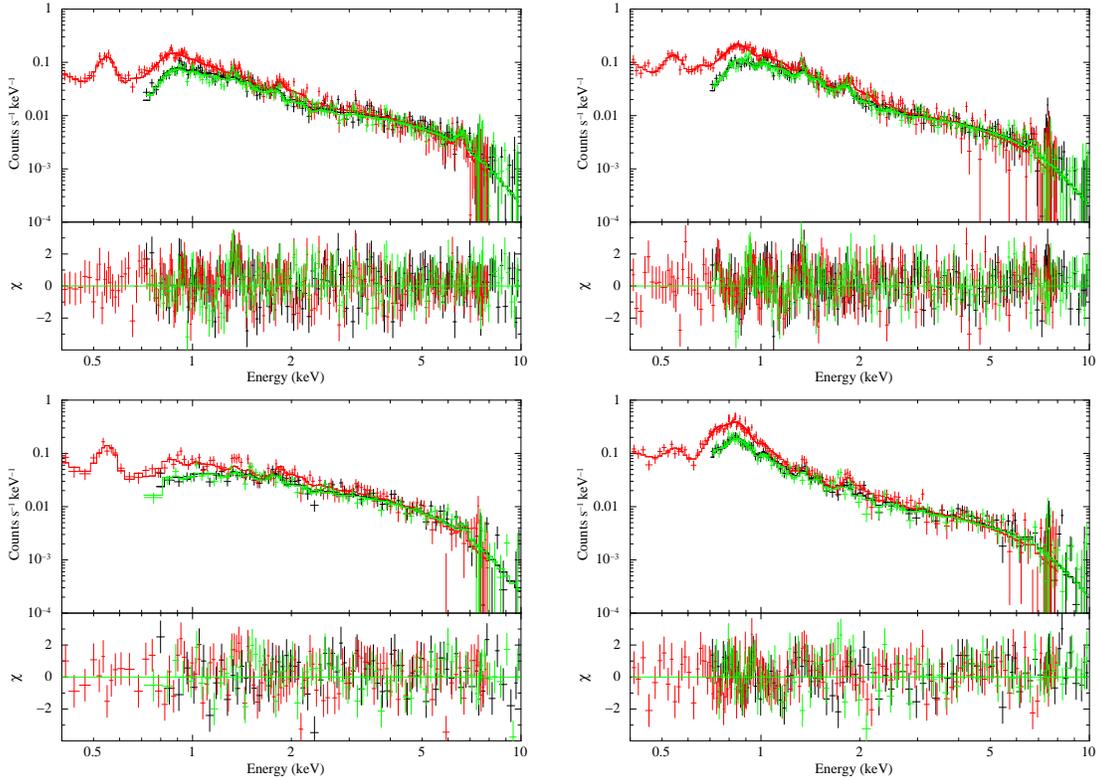

\centering
\includegraphics[height=0.45\textwidth,angle=-90]{S1_xi013_diffuse_base3_fit.cps}
\includegraphics[height=0.45\textwidth,angle=-90]{S2_xi013_diffuse_base3_fit.cps}
\includegraphics[height=0.45\textwidth,angle=-90]{BG1_xi013_diffuse_base3_fit.cps}
\includegraphics[height=0.45\textwidth,angle=-90]{BG2_xi013_diffuse_base3_fit.cps}
\caption{X-ray spectra of extended emissions integrated over the XIS field-of-view of 
Source~1 observation (top left), Source~2 observation (top right), Background~1 observation (bottom left), 
and Background~2 observation (bottom right).
The XIS0, 1, and 3 data are indicated by black, red, and green crosses, respectively. 
Solid lines are the best-fit models. The bottom panels show the residuals.
}
\end{figure}

\subsection{Spectra of Individual Point Sources and Small-scale Structures}

To study point sources and small-scale structures, 
we have integrated the events within the $1.\hspace{-2pt}'5$ radius circular 
regions 
shown in Figure~2. If the $1.\hspace{-2pt}'5$ radius was too large and contaminated from nearby sources 
(e.g., src1 and src2 in Source 1 observation) or if the source was close to the edge of the CCD chip 
(e.g., src6 in Source~1 observation), the radius has been reduced to $1'$. 
The RMF and ARF have been calculated by {\tt xisrmfgen} and {\tt xisarfgen}, respectively, 
assuming a point-source response in the latter. 
For the spectral fitting, we have used an absorbed power-law model 
(${\tt wabs} \times {\tt pow}$ in {\tt XSPEC}) or an absorbed thin-thermal plasma model
(${\tt wabs} \times {\tt apec}$ in {\tt XSPEC}).
We have first tried a power-law model and employed a thin-thermal plasma model 
if the power-law model was inadequate; i.e., if we observed large residuals in the fit
or if the best-fit photon index were excessively large ($\Gamma \ge 3$).
Src3 in Source~1 observation was not well fitted 
by a single component model, so we have 
fitted its spectrum using a two-component model 
({\tt apec}+{\tt pow} with the common absorption).
The fits results are summarized in Table~3. The majority of sources have a 
hard power-law spectrum 
with a photon index $\Gamma$ of $\le 2.5$ and relatively large absorption 
($N(\mathrm H) \ge 2 \times 10^{22}~{\rm cm^{-2}}$).
These sources (src2, src4 and src5 in Source~1 observation, src1 and src2 in Source~2 observation,
src1 in Background~1 observation and src1 in Background~2 observation) 
are most likely background active-galactic nuclei. 
According to \citet[][]{Tozzi2001} who compiled the $\log N\mbox{--}\log S$ relation of X-ray sources,
the expected number of active-galactic nuclei with 2--10~keV flux $\ge 0.5 \times 10^{-13}~{\rm erg~s^{-1}~cm^{-2}}$
is ${\rm \sim 30/deg^{2}}$. This implies 2--3 uncatalogued active-galactic nuclei in each
XIS field-of-view, which is close to the number of sources we found.
Possible counterparts, identified from the 
SIMBAD\footnote{\url{http://simbad.u-strasbg.fr/simbad/sim-fcoo}} and
NED\footnote{\url{http://ned.ipac.caltech.edu/}}
databases within $1'$ of the source position, are also listed in Table~3.

\begin{landscape} 
\begin{table}
\begin{center}
\caption{Spectra of point sources and small-scale structures}
\tiny
\begin{tabular}{ccccccccc}
 \tableline\tableline
Region/source & $N_{\mathrm{H}}$ & $\Gamma$ & $kT$ & $A$        & $f_{0.5\mbox{--}2}$        & $f_{2\mbox{--}10}$        & $\chi^{2}$/d.o.f. & possible \\ 
       & $10^{22}~{\rm cm^{-2}}$ &          & keV  & $Z_{\sun}$ & ${\rm erg~s^{-1}~cm^{-2}}$ & ${\rm erg~s^{-1}~cm^{-2}}$ & & counterpart\tablenotemark{a} \\ 
\tableline
Source 1 & & & & & & & & \\
\hspace{0.15cm} src1 & $0.03(\le 0.09)$ & -- & $0.94^{+0.08}_{-0.14}$ & $0.07 \pm 0.03$ & 
$1.81 \times 10^{-13}$ & $0.29 \times 10^{-13}$ & 71.6/64 & 1RXS J202725.2+405428\\ 
\hspace{0.15cm} src2 & $5.1^{+5.2}_{-2.7}$ & $2.3^{+1.5}_{-1.0}$ & -- & -- &
$0.03 \times 10^{-13}$ & $1.52 \times 10^{-13}$ & 41.6/40 & NVSS J202655+405408 \\ 
\hspace{0.15cm} src3 & $0.79^{+0.59}_{-0.45}$ & $-1.0(\ge -2.2)$ & $0.55^{+0.48}_{-0.36}$ & 1(fixed) &
$0.19 \times 10^{-13}$ & $1.27 \times 10^{-13}$ & 28.1/25 & NVSS J202723+405706 \\ 
\hspace{0.15cm} src4 & 2.2($\le 6.7$) & $1.5^{+1.6}_{-1.3}$ & -- & -- &
$0.06 \times 10^{-13}$ & $1.47 \times 10^{-13}$ & 16.0/25 & TYC 3156-1302-1 \\ 
\hspace{0.15cm} src5 & $9^{+15}_{-7}$ & 1.6($\le 4.1$) & -- & -- &
$\le 0.01 \times 10^{-13}$ & $1.75 \times 10^{-13}$ & 32.1/26 & NVSS J202642+405138 \\ 
\hspace{0.15cm} src6 & 0($\le 0.07$) & $1.72^{+0.47}_{-0.43}$  & -- & -- &
$0.48 \times 10^{-13}$ & $0.86 \times 10^{-13}$ & 18.1/13 & --\\ 
Source 2 & & & & & & \\
\hspace{0.15cm} src1 & $2.33^{+0.86}_{-0.69}$ & $1.58^{+0.34}_{-0.30}$ & -- & -- & 
$0.24 \times 10^{-13}$ & $5.46 \times 10^{-13}$ & 120.8/101 & --\\ 
\hspace{0.15cm} src2 & 2.47(fixed\tablenotemark{b}) & $1.23^{+0.73}_{-0.83}$ & -- & -- & 
$0.02 \times 10^{-13}$ & $0.92 \times 10^{-13}$ & 20.3/24 & --\\ 
\hspace{0.15cm} src3 & $0.36^{+0.42}_{-0.30}$ & $1.92^{+0.49}_{-0.41}$ & -- & -- & 
$0.32 \times 10^{-13}$ & $0.91 \times 10^{-13}$ & 81.7/58 & --\\ 
Background 1 & & & & & & \\
\hspace{0.15cm} src1 & $2.5^{+1.7}_{-1.3}$ & $2.6^{+1.0}_{-0.9}$ & -- & -- & 
$0.34 \times 10^{-13}$ & $3.00 \times 10^{-13}$ & 18.0/19 & several\tablenotemark{c} \\ 
\hspace{0.15cm} src2 & $3.1^{+8.8}_{-1.8}$ & -- & $0.05^{+2.05}_{-0.02}$ & 1(fixed) & 
$0.47 \times 10^{-13}$ & $\le 0.01 \times 10^{-13}$ & 1.1/1 & --\\ 
Background 2 & & & & & & \\
\hspace{0.15cm} src1 & $0(\le 18)$ & $1.0(\le 5.4)$ & -- & -- & 
$0.15 \times 10^{-13}$ & $0.79 \times 10^{-13}$ & 15.0/17 & NVSS J202754+423205 \\ 
\hspace{0.15cm} src2 & $1.5^{+3.7}_{-1.3}$ & -- & $0.17^{+0.89}_{-0.14}$  & 1(fixed) & 
$0.13 \times 10^{-13}$ & $\le 0.01 \times 10^{-13}$ & 17.5/23 & TYC 3160-1261-1 \\ 
\tableline
\end{tabular}
\tablenotetext{}{$\Gamma$ is the photon index of the power-law model and $kT$ and $A$ are the temperature and abundance, respectively, in the plasma model.
Abundance was fixed at the solar value if the value is not well constrained by a fit.
$f_{0.5\mbox{--}2}$ and $f_{2\mbox{--}10}$ are absorption-uncorrected fluxes in 
0.5--2~keV and 2--10~keV, respectively.}
\tablenotetext{a}{Objects named NVSS and TYC are radio sources \citep{Condon1998} and stars \citep{Hog1998}, respectively. IRXS denotes a source detected in the ROSAT All-sky survey \citep{Voges1999}.
}
\tablenotetext{b}{$N({\mathrm H})$ is not well determined and is fixed to the estimated Galactic absorption.}
\tablenotetext{c}{Crowded region with several possible counterparts (stars, radio sources and
{{\sc H}\,{\scriptsize{\sc II}}} regions.)}
\end{center}
\end{table}
\end{landscape} 

\clearpage

\section{Discussion}

\subsection{Constraint on X-Ray Emission from the Cocoon}

As shown in Table~2 and Figure~4, no significant excess has been found in the X-ray spectra 
of the extended emission in any of the four regions. 
However, the data have been fitted with the CXB+GRXE(plus local diffuse X-ray emission)
model described in \S~3.2 and the fitted GRXE component may include part of the cocoon emission, if any. 
To search for an excess above the expected GRXE profile, we have 
examined the positional dependence 
of the CXB-subtracted intensity. We have focused on the hard-band (2--10~keV), 
in which the relative contribution of an extended X-ray emission from the cocoon 
is expected to be highest.

Figure~5 shows how the intensity varies with Galactic latitude. 
The intensity of the CXB contribution is somewhat uncertain, due to uncertainty in the absorption 
and the field-to-field fluctuations of the CXB. 
To gauge the uncertainty of the absorption, we repeated the calculation of
$N(\mathrm H)$ with spin temperature of 100~K and optical-thin approximation,
and found the impact on 2--10~keV intensity is $\le 0.46 \times 10^{-8}~{\rm erg~s^{-1}~cm^{-2}~sr^{-1}}$.
According to \cite{Kushino2002}, the field-to-field fluctuation of the CXB in the 
{\it ASCA}-GIS field-of-view ($\sim 0.4~{\rm deg^{2}}$) is 5.1\% ($1 \sigma$).
Assuming that the fluctuation is Poisson's noise, 
the expected fluctuation over the entire XIS field-of-view 
($\sim 17.\hspace{-2pt}'8 \times 17.\hspace{-2pt}'8$) is $5.1 \times \sqrt{0.4/0.088} = 11\%$.
Therefore the uncertainty of the CXB contribution in the 2--10~keV range is
$(0.52\mbox{--}0.54) \times 10^{-8}~{\rm erg~s^{-1}~cm^{-2}~sr^{-1}}$.
Uncertainty in the NXB model also affects the intensity of the extended emission. 
\cite{Tawa2008} reported that the reproducibility of the XIS NXB model ($1 \sigma$) is 
$\sim 5\%$ and 3--4\% in 1--7~keV and 5--12~keV ranges, respectively.
We raised and lowered the NXB model by 5\%, and found that the NXB-subtracted intensity 
in 2--10~keV was altered by $(0.26\mbox{--}0.63)\times 10^{-8}~{\rm erg~s^{-1}~cm^{-2}~sr^{-1}}$.
We added the uncertainties due to the fluctuation of the CXB intensity and the NXB model reproducibility
in quadrature, and took the effect from the assumption of spin temperature as systematic uncertainty.
Even allowing for these uncertainties, the intensity profile (Figure~5) shows a clear monotonic decrease 
with increasing latitude, as expected for the GRXE. 
We have therefore concluded that most of the extended emission in the CXB-subtracted spectra 
in 2--10~keV is contributed by GRXE.

We therefore impose constraints on the extended X-ray emission from the cocoon. 
Among the four positions, the Galactic latitude is the highest in Background~2. 
Therefore, we expect the GRXE intensity to be minimized there and take the intensity of Background 2 
as the lower limit of the GRXE contribution in the source positions. 
A robust upper limit on the extended X-ray emission from the cocoon can be 
obtained 
by subtracting the intensity in Background~2. In this way, 
the upper limit in the 2--10~keV range for Source~1 and Source~2 was determined as 
$3.4 \times 10^{-8}~{\rm erg~s^{-1}~cm^{-2}~sr^{-1}}$ and
$1.3 \times 10^{-8}~{\rm erg~s^{-1}~cm^{-2}~sr^{-1}}$, respectively. 

We have also calculated the expected synchrotron X-ray intensity from a CR electron spectrum which can reproduce
the GeV emission of the cocoon by up scattering the interstellar and stellar radiation fields modelled in the region \citep{Ackermann2011}. The latter include the bright stars of the Cyg OB2 and NGC 6910 associations, an average spectrum for the field stars, the intense IR field from the heated dust in the region, and the cosmological microwave background. The electron spectrum is based on the LIS one, which cuts off near 1~TeV, with power-law extensions up to 5, 10, 50, or 100~TeV. These spectra being too soft, we have applied an
amplification factor of ${\rm 120 \times (E/50~GeV)^{0.5}}$ to account for the \textit{Fermi}-LAT data. It yields an electron density of $2.6 \times 10^{-9}~{\rm m^{-3}~TeV^{-1}}$ at 1~TeV. We have assumed a thickness of 25~pc for the emission region along the line-of-sight (corresponding to $1^{\circ}$ at 
$d=1.4~{\rm kpc}$ in the plane of the sky), and a
magnetic field strength of ${\rm 20~\mu G}$ 
(deduced from pressure balance between the magnetic field and the 
gas)\footnote{
Although this value is larger than the local interstellar magnetic field 
strength, we have adopted it as \citet[][]{Ackermann2011} examined possible 
solutions to explain the GeV/TeV $\gamma$-ray data while not violating the
radio data in the specific environment of the cocoon.}. 
The synchrotron intensity expected in the 2--10~keV 
range from the electron spectrum extended to 100~TeV is
$\sim 7.0 \times 10^{-8}~{\rm erg~s^{-1}~cm^{-2}~sr^{-1}}$.
As the upper limits from the {\it Suzaku}-XIS observations are 1/5--1/2 of this value, 
we infer that our data are inconsistent with a pure inverse-Compton scenario in $\gamma$ rays without a 
cut off in the electron spectrum.

To further constrain the CR properties, we have investigated the multi-wavelength spectra, 
including radio and TeV $\gamma$-rays as summarized in Figure~\ref{fig_MWspec}.
To compensate for the small field-of-view of the X-ray data compared to the size of the cocoon,
we have integrated the averaged X-ray upper-limit 
(2--10~keV intensity of $2.35 \times 10^{-8}~{\rm erg~s^{-1}~cm^{-2}~sr^{-1}}$, 
or $9.13 \times 10^{-3}~{\rm MeV^{2}~s^{-1}~cm^{-2}~sr^{-1}~MeV^{-1}}$ assuming a photon index of 2.0)
over the solid angle subtended by the 
cocoon \citep[${\rm 4.38 \times 10^{-3}~sr}$;][]{Ackermann2011}.
A potential TeV counterpart to the cocoon, MGRO~J2031+41, has been found by MILAGRO \citep[][]{Abdo2007,Abdo2012}. 
More recent results from the ARGO-YBJ experiment 
\citep{Bartoli2014} confirm the presence of 
an extended TeV source of compatible position and size, even after subtracting the contribution
from the known overlapping/nearby TeV sources. Its spectrum smoothly connects to
the {\it Fermi}-LAT data points as described by \citet{Bartoli2014}
(see also Fig.~\ref{fig_MWspec})\footnote{
Although VERITAS has surveyed the cocoon region, it was not able to detect the
$\gamma$-ray emission on the angular scale of the Cygnus cocoon and MGRO~J2031+41
with a standard ring-background estimation method, as described by \cite{Aliu2013}.
}.
The radio flux has been averaged over the cocoon from the CGPS survey \citep{Taylor2003}.
The values should be considered as upper limits to the possible synchrotron emission from the cocoon because free-free emission from the hot ionized gas of Cygnus X has not been removed. The WMAP free-free map is too coarse and H$\alpha$ intensities are too absorbed in this region to allow a reliable subtraction of the free-free emission. 
The plotted curves for the synchrotron and inverse Compton models assume the amplified CR electron spectra described above, with the LIS cutting off near 1~TeV, and with power-law extensions to 5~TeV, 10~TeV, 50~TeV, and 100~TeV cut-offs.
Our X-ray data require a cut-off in electron energy
below ${\rm 50~TeV}$, in agreement with the marked decline in TeV emission seen above 2~TeV \citep{Abdo2012,Bartoli2014}. 
The X-ray data should therefore be regarded as an independent confirmation of 
the spectral cutoff in the electron 
scenario inferred from TeV observations.

\subsection{Properties of GRXE and Local Diffuse X-rays}

Finally we comment on the properties of GRXE and local diffuse X-ray emission.
All four observations exhibit emission lines in the 0.5--0.6~keV range (see 
Figure~4). As shown in Table~2, the
line central energy is ${\rm \sim 0.55~keV}$ and is not compatible with 
the {{\sc O}\,{\scriptsize{\sc VII}}}
resonance line energy (0.574~keV) expected for the local hot plasma,
but it is marginally consistent with the {{\sc O}\,{\scriptsize{\sc VII}}} forbidden line energy (0.561~keV)
which is a signature of a charge-exchange process
\citep[e.g.,][]{Ebisawa2008,Fujimoto2007}.
We therefore conclude that SWCX significantly contributes to our data below 1~keV. 

The intensity of the mid-temperature component ($I_{\rm mid}$ in Table~2) shows a
positive correlation with Galactic latitude.
This unusual behavior for GRXE may be attributed to the structure of the Cygnus Super Bubble (CSB).
As shown by \citet[][]{Uyaniker2001}, the positions of the Source~2 and Background~2 
images are inside or
very close to the rim of the CSB, as seen by ROSAT in the 0.75 and 1.5~keV bands. Hence the intensity
of the mid-temperature
component should be significantly increased.
We note that the $N({\mathrm H})_{\rm mid}$ and $kT_{\rm mid}$ values
of Source~2 and Background~2 are not compatible
with the $N({\mathrm H}) \sim 0.3 \times 10^{22}~{\rm cm^{-2}}$ and $kT \sim 0.3~{\rm keV}$ values
ubiquitously found in CSB by \citet[][]{Kimura2013}. In particular, the Background~2 spectrum
shows strong 0.7--1~keV emission (Fe-L and Ne lines),
but lacks the line of {{\sc O}\,{\scriptsize{\sc VIII}}},
and therefore is not compatible
with plasma emission at $\sim$ 0.3~keV.
Although {\it Suzaku} provides high-quality spectra compared to those in \citet{Kimura2013}
in terms of energy range and resolution, our data correspond to a much smaller region, so a
detailed discussion of the origin of this discrepancy is difficult and beyond the
scope of this study.

The latitude dependence of the high-temperature component ($I_{\rm high}$ in Table~2)
gives a scale height of $\sim 1.\hspace{-2pt}^{\circ}5$, which is 
2--3 times larger than the GRXE scale height in the Scutum arm region close to 
the Galactic plane
\citep[$\sim 0.\hspace{-2pt}^{\circ}5$; e.g.,][]{Kaneda1997}.
This difference may be attributed to the proximity of Cygnus X (distance $\sim 
1.4~{\rm kpc}$), or to the different Galactic latitude 
ranges of the two studies
($b = 0.7\mbox{--}2.2$~deg in our observations whereas \citet[][]{Kaneda1997} analyzed data
in $b = 0\mbox{--}1.7$~deg).
The observed temperature (2--3~keV) is significantly lower than
the value of hard-temperature component of the GRXE reported for lower Galactic longitudes;
e.g., $kT$ of $\sim 10~{\rm keV}$ and $\sim 5~{\rm keV}$ were reported at 
$l \sim 28^{\circ}$ \citep[][]{Kaneda1997} and 
$l \sim 18^{\circ}$ \citep[][]{Uchiyama2009b}, respectively.
These Galactic longitude dependence of the temperature may provide 
useful information
on the origin of GRXE (e.g., different populations of sources if GXRE is due 
to unresolved sources),
but we defer a detailed discussion on this topic.

\begin{figure}[htbp]
\centering
\includegraphics[width=0.6\textwidth]
{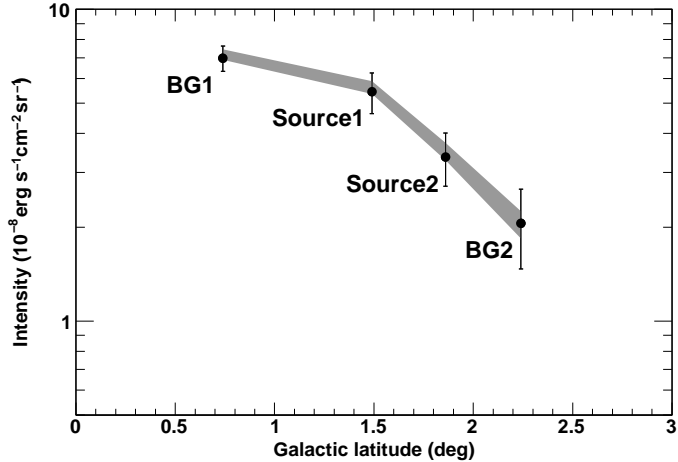}
\caption{Dependence of the hard-band intensity (2--10~keV) on the
Galactic latitude after subtracting the CXB contribution. 
The field-to-field fluctuation of the CXB intensity and the NXB model reproducibility
are indicated by error bars, and 
the systematic uncertainty due to the absorption
of the CXB model is shown by a shaded area.}
\end{figure}

\begin{figure}[htbp]
\centering
\includegraphics[width=0.8\textwidth]{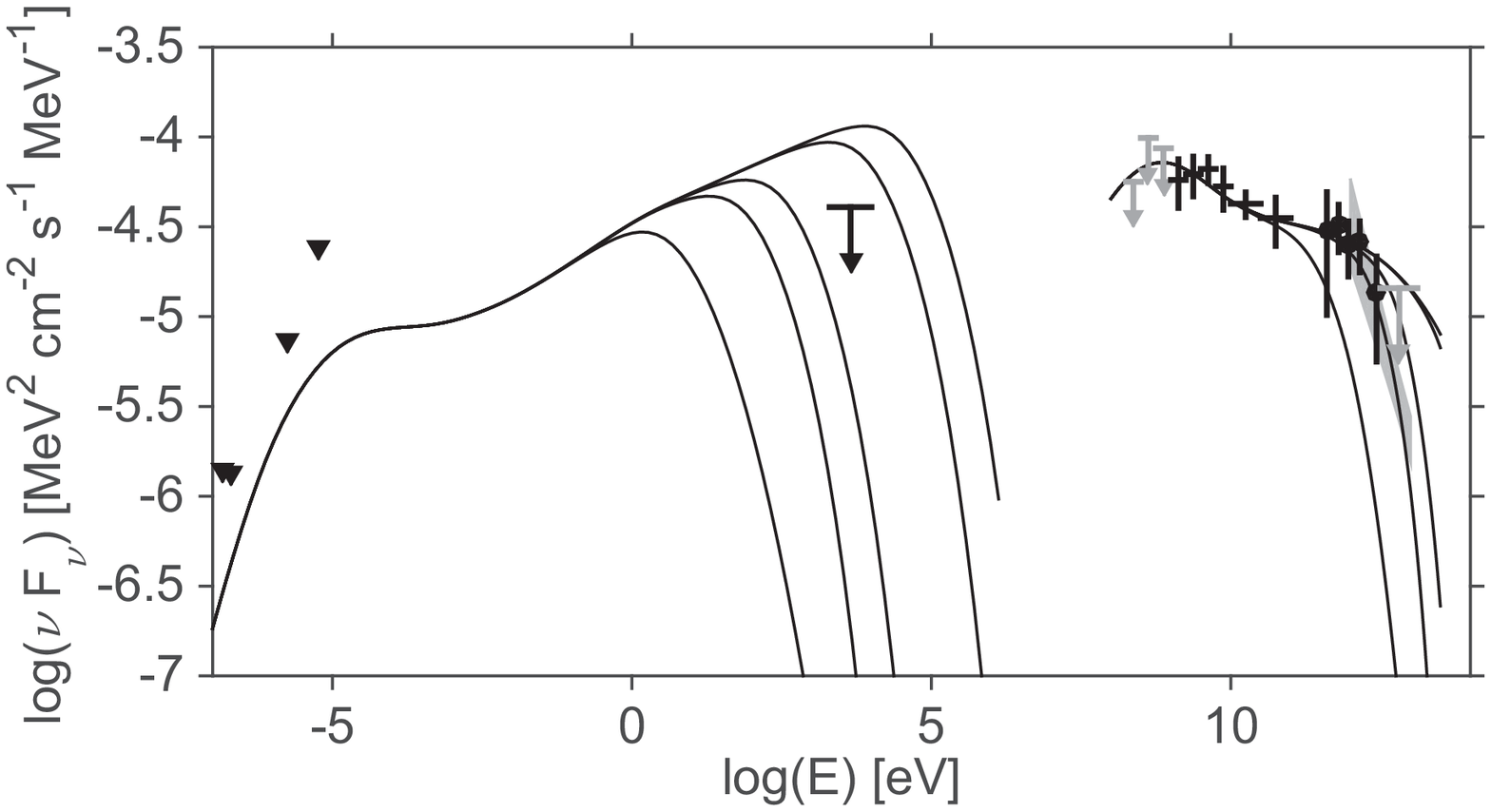}
\caption{Multi-wavelength data and models of the emission from the Cygnus cocoon region: 
the data points in the 0.1--100 GeV band come from the {\it Fermi}-LAT data \citep{Ackermann2011}, 
the data points and the power-law model at TeV energies from the ARGO-YBJ \citep{Bartoli2014} and
the MILAGRO \citep{Abdo2012} experiments, respectively,
the X-ray upper limit from the present work, and the radio upper limits (triangles) from the CGPS survey \citep{Taylor2003}. 
At TeV energies the MILAGRO data likely include all sources around the cocoon area, 
but below 10~TeV the flux exceeds the sum from TeV J2032+4130 \citep{Aharonian2005}
and VER J2019+407 \citep{Aliu2013} by at least a factor of two.
Thin lines show the modelled 
synchrotron and inverse Compton emissions, assuming an amplified LIS electron spectrum, or this spectrum with power-law extensions to cut-off energies of 5, 10, 50, and 100~TeV.
See text for details.
}
\label{fig_MWspec}
\end{figure}

\section{Summary}

We have conducted a series of deep X-ray observations of the nearby star-forming region of Cygnus X
using {\it Suzaku} XIS, in order to better understand the origin of the
GeV emission revealed by {\it Fermi}-LAT in the cocoon of young cosmic rays pervading Cygnus X. 
After excluding point sources and small-scale structures from the X-ray images, and after subtracting the CXB, 
we find that
the X-ray distribution in 2--10~keV band monotonically decreases with increasing 
Galactic latitude, thus
indicating that most of the extended emission relates to the GRXE. 
We have obtained robust upper limits to the diffuse X-ray 
emission from the $\gamma$-ray cocoon.
The results are incompatible with a pure inverse-Compton scenario for the 
origin of the $\gamma$-ray emission without a cut-off in the electron spectrum 
above about ${\rm 50~TeV}$, in agreement with the drop observed in the $\gamma$-ray spectrum above 2~TeV. 
We have found a strong contribution of SWCX below 1~keV in our data.
The mid-temperature component 
($kT=0.6\mbox{--}0.7~{\rm keV}$)
is likely to be significantly affected by the rim of CSB.
The scale-height of the high-temperature component intensity ($\sim 1.\hspace{-2pt}^{\circ}5$) is 
2--3 times larger than the scale height of GRXE in the Scutum arm region at $l 
\sim 28^{\circ}$.
We attribute this increased scale height to the proximity of Cygnus X.
The observed temperature of the high-temperature component is 2--3~keV and is 
lower than found for the GRXE
at smaller Galactic longitudes. 

We would like to thank K. Ebisawa, K. Hayashida and S. Yamauchi for valuable comments.
We also thank the Suzaku team members for their dedicated support of the satellite operation
and calibration.

\clearpage





\begin{thebibliography}{}

\bibitem[Abdo et al.(2007)]{Abdo2007}
Abdo, A.~A., Allen, B., Berley, D., et al. 2007, \apjl, 664, 91

\bibitem[Abdo et al.(2012)]{Abdo2012}
Abdo, A.~A., Abeysekara, A.~U., Allen, B.~T., et al. 2012, \apj, 753, 159

\bibitem[Ackermann et al.(2011)]{Ackermann2011}
Ackermann, M., Ajello, M., Allafort, A., et al. 2011, Science, 334, 1103

\bibitem[Ackermann et al.(2012a)]{Paper2}
Ackermann, M., Ajello, M., Atwood, W.~B., et al. 2012a, \apj, 750, 3

\bibitem[Ackermann et al.(2012b)]{CygISM}
Ackermann, M., Ajello, M., Allafort, A., et al. 2012b, A\&A, 538, 71

\bibitem[Atwood et al.(2009)]{Atwood2009}
Atwood, W.~B., Abdo, A.~A., Ackermann, M., et al. 2009, \apj, 697, 1071

\bibitem[Aharonian et al.(2005)]{Aharonian2005}
Aharonian, F., Akhperjanian, A., Beilicke, M., et al. 2005, A\&A, 431,197

\bibitem[Aliu et al.(2013)]{Aliu2013}
Aliu, E., Archambault, S., Arlen, T., et al. 2013, \apj, 770, 93

\bibitem[Bartoli et al.(2014)]{Bartoli2014}
Bartoli, B., Bernardini, P., Bi, X.~J., et al. 2014, \apj, 790, 152

\bibitem[Bykov \& Fleishman(1992)]{bykov92} 
Bykov, A.~M., \& Fleishman, G.~D.\ 1992, \mnras, 255, 269

\bibitem[Condon et al.(1998)]{Condon1998}
Condon, J.~J., Cotton, W.~D., Greisen, E.~W., et al. 1998, AJ, 115, 1693

\bibitem[Ebisawa et al.(2005)]{Ebisawa2005}
Ebisawa, K., Tsujimoto, M., Paizis, A., et al. 2005, \apj, 635, 214

\bibitem[Ebisawa et al.(2008)]{Ebisawa2008}
Ebisawa, K., Yamauchi, S., Tanaka, Y., et al. 2008, PASJ, 60, S223

\bibitem[Fujimoto et al.(2007)]{Fujimoto2007}
Fujimoto, R., Mitsuda, K., Mccammon, D., et al. 2007, PASJ, 59, 133

\bibitem[Ginzburg \& Syrovatskii(1964)]{Ginzburg1964}
Ginzburg V.I., \& Syrovatskii S.I. 1964, {\it the Origin of Cosmic Rays}, Pergamon Press. Classic monograph

\bibitem[Grenier et al.(2005)]{DNM}
Grenier, I.~A., Casandjian, J.-M., Terrier, R. 2005, Science 307, 1292

\bibitem[Higdon \& Lingenfelter(2005)]{higdon05} Higdon, J.~C., \& Lingenfelter, R.~E.\ 2005, \apj, 628, 738 

\bibitem[Hog et al.(1998)]{Hog1998}
Hog, E., Kuzmin, A., Bastian, U., et al., 1998, A\&A, 335, 65

\bibitem[Ishisaki et al.(2007)]{Ishisaki2007}
Ishisaki, Y., Maeda, Y., Fujimoto, R., et al. 2007, PASJ, 59, S113

\bibitem[Kalberla et al.(2005)]{Kalberla2005}
Kalberla, P.~M.~W., Burton, W.~B., Hartmann, D., et al. 2005, \aap, 440, 775

\bibitem[Kaneda et al.(1997)]{Kaneda1997}
Kaneda, H., Makishima, K., Yamauchi, S., et al. 1997, \apj, 491, 638

\bibitem[Kataoka et al.(2013)]{Kataoka2013}
Kataoka, J., Tahara, M., Totani, T., et al. 2013, ApJ 779, 57

\bibitem[Kimura et al.(2013)]{Kimura2013}
Kimura, M., Tsunemi, H., Tomida, H., et al. 2013, PASJ, 65, 14

\bibitem[Koyama et al.(1986)]{Koyama1986}
Koyama, K., Makishima, K., Tanaka, Y., \& Tsunemi, H. 1986, PASJ, 38, 121

\bibitem[Koyama et al.(2007)]{Koyama2007}
Koyama, K., Tsunemi, H., Dotani, T., et al. 2007, PASJ, 59, S23

\bibitem[Kushino et al.(2002)]{Kushino2002}
Kushino, A., Ishisaki, Y., Morita, U., et al. 2002, PASJ, 54, 327

\bibitem[Mitsuda et al.(2007)]{Mitsuda2007}
Mitsuda, K., Bautz, M., Inoue, H., et al. 2007, PASJ, 59, S1

\bibitem[Morrison \& MCammon (1983)]{Morrison1983}
Morisson, R., \& McCammon, D. \apj, 270, 119

\bibitem[Parizot et al.(2004)]{parizot04} 
Parizot, E., Marcowith, A., van der Swaluw, E., et al.\ 2004, \aap, 424, 747 

\bibitem[Piddington \& Minnett(1952)]{Piddington1952}
Piddington, J.~H., \& Minnett, H.~C. 1952, Australian J. Sci. Res., 5, 17

\bibitem[Revnivtsev et al.(2006)]{Revnivtsev2006}
Revnivtsev, M., Sazonov S., Gilfanov, M., Churazov, E., \& Sunyaev, E. 2006, A\&A, 452, 169

\bibitem[Rygl et al.(2012)]{rygl12} 
Rygl, K.~L.~J., Brunthaler, A., Sanna, A., et al.\ 2012, \aap, 539, 79 

\bibitem[Serlemitsos et al.(2007)]{Serlemitsos2007}
Serlemitsos, P.~J., Soong, Y., Chan, K.-W., et al. 2007, PASJ, 59, 59

\bibitem[Smith et al.(2007)]{Smith2007}
Smith, R.~K., Bautz, M.~W., Edgar, R.~J., et al. 2007, PASJ 59, S141

\bibitem[Smith et al.(2001)]{Smith2001}
Smith, R.~K., Brickhouse, N.~S., Liedahl, D.~A., \& Raymond, J.~C. 2001, \apjl, 556, 91

\bibitem[Tawa et al.(2008)]{Tawa2008}
Tawa, N., Hayashida, K., Nagami, M., et al. 2008, PASJ, 60, 11

\bibitem[Taylor et al.(2003)]{Taylor2003}
Taylor, A.~R., Gibson, S.~J., Peracaula, M., et al. 2003, \aj, 125, 3145

\bibitem[Tenorio-Tagle \& Bodenheimer(1988)]{tenorio88} 
Tenorio-Tagle, G., \& Bodenheimer, P.\ 1988, \araa, 26, 145 

\bibitem[Tozzi et al.(2001)]{Tozzi2001}
Tozzi, P., Rosati, P., Nonino, M., et al. 2001, \apj, 562, 42

\bibitem[Uchiyama et al.(2009a)]{Uchiyama2009a}
Uchiyama, H., Ozawa, M., Matsumoto, H. et al. 2009a, PASJ, 61, S9

\bibitem[Uchiyama et al.(2009b)]{Uchiyama2009b}
Uchiyama, H., Matsumoto, H., Tsuru, T.~G., Koyama, K., \& Bamba, A. 2009b, PASJ, 61, S189

\bibitem[Uchiyama et al.(2013)]{Uchiyama2013}
Uchiyama, H., Nobukawa, M., Tsuru, T.~G., \& Koyama, K. 2013, PASJ, 65, 19

\bibitem[Uyaniker et al.(2001)]{Uyaniker2001}
Uyaniker, H., F\"{u}rst, E., Reich, W., Aschenbach, B., \& Wielebinski, R. 2001, A\&A, 371, 675

\bibitem[Voges et al.(1999)]{Voges1999}
Voges, W., Aschenbach, B., Boller, T., et al. 1999, A\&A, 349, 389

\bibitem[Warwick et al.(1985)]{Warwick1985}
Warwick, R.~S., Turner, M.~J.~L., Watson, M.~G., \& Willingale, R. 1985, Nature, 317, 218

\bibitem[Worrall et al.(1982)]{Worrall1982}
Worrall, D.~M., Marshall, F.~E., Boldt, E.~A., \& Swank, J.,~M. 1982, \apj, 255, 111

\bibitem[Yoshino et al.(2009)]{Yoshino2009}
Yoshino, T., Mitsuda, K., Yamasaki, N.~Y., et al. 2009, PASJ, 61, 805

\end{thebibliography}
\end{document}